\documentclass[prd,nofootinbib,showpacs,preprint]{revtex4}

\usepackage{graphicx}
\usepackage[usenames,dvipsnames]{color}
\usepackage{amsmath,amssymb}
\usepackage[dvipdfm,colorlinks,citecolor=blue]{hyperref}
\begin{document}
\title{Common Radiative Origin of Active and Sterile Neutrino Masses}
\author{Debasish Borah}
\email{dborah@tezu.ernet.in}
\affiliation{Department of Physics, Tezpur University, Tezpur - 784028, India}
\author{Rathin Adhikari}
\email{rathin@ctp-jamia.res.in}
\affiliation{Centre for Theoretical Physics, Jamia Millia Islamia - Central University, Jamia Nagar, New Delhi - 110025, India}
\begin{abstract}
Sterile neutrinos with sub-electron volt (eV) masses have recently received serious attention due to the tantalizing hints from reactor neutrino experiments as well as cosmology. While the nine year old Wilkinson Mass Anisotropy Probe experiment suggests the effective number of relativistic degrees of freedom to be $N_{\text{eff}} = 3.84 \pm 0.40$, recently reported Planck collaboration results show more preference towards the standard three light neutrino scenario $ N_{\text{eff}} = 3.30^{+0.54}_{-0.51}$. Keeping in mind that the issue of existence or non-existence of sub-eV scale sterile neutrinos is not yet settled, here we outline a mechanism to generate sub-eV scale masses for three active and one sterile neutrinos simultaneously. The model is based on an abelian extension of standard model where the fermion and scalar fields are charged under the additional $U(1)$ gauge group in such an anomaly free way that it allows one eV scale neutrino and three massless neutrinos at tree level. However, at one loop level, this model naturally allows three active and one sterile neutrino with mass at the sub-eV scale. The model also allows for mixing between active and sterile neutrinos at one loop level which can have interesting signatures in reactor neutrino experiments.
\end{abstract}
\maketitle

\section{Introduction}

The Standard Model (SM) of particle physics has turned out to be the most successful low energy theory, specially after the 2012 discovery of its last missing piece: the Higgs boson. Despite its phenomenological success, the standard model neither addresses 
some theoretical issues like 
gauge hierarchy problem, nor provides a complete understanding of various observed phenomena like non-zero neutrino masses, dark matter etc. A significant amount of works have been carried out so far on various possible extensions of the Standard model, although none of them can be called a complete 
phenomenological model. Such extensions usually involve incorporating some additional symmetries (gauged or global) into the Standard model or inclusion of additional fields. We know that the smallness of three Standard Model
neutrino masses \cite{neutosc,noscpheno} can be naturally explained 
via seesaw mechanism. Such seesaw mechanism can be of three types : type I \cite{typeI}, type II \cite{typeII} and type III \cite{Foot:1988aq}. All these mechanisms involve the inclusion of additional fermionic or scalar fields to generate tiny neutrino masses at tree level. However, it could well be true that the gauge symmetry as well as the field content of the theory do not allow neutrino masses at tree level and tiny neutrino masses appear only at the loop level. Here we are interested in a model which gives rise to such radiative neutrino mass in the manner proposed in \cite{Ma:2006km,Adhikari:2008uc}.

In addition to additional scalar and fermionic fields, the model we study also has an enhanced gauge symmetry: an additional $U(1)_X$ gauge symmetry. It is worth mentioning that abelian gauge extension of Standard Model is one of the best motivating examples of beyond Standard Model physics \cite{Langacker:2008yv}. Such a model is also motivated within the framework of GUT models, for example $E_6$. 
The supersymmetric version of such models can also provide a solution to the MSSM $\mu$ problem, among many other advantages. An abelian gauge extension of SM was studied recently by one of us in the context of four fermion generations \cite{Borah:2011ve} which explains the origin of three light and one heavy fourth generation neutrino masses and at the same time provides a way to avoid the strict bounds put by Large Hadron Collider (LHC) on a SM like Higgs boson mass in the presence of a fourth family.

Recently, a similar abelian gauge extension of Standard Model was studied in the context of radiative neutrino mass and dark matter in \cite{Borah:2012qr}. Here we study the same model with little modification to take into account light sterile neutrinos. Light sterile neutrino of mass of the order of electron volts (eV) have recently got lots of attention due to some experimental evidence suggesting additional light degrees of freedom beyond the three active neutrino species. For a review, please see \cite{whitepaper}. The nine year Wilkinson Mass Anisotropy Probe (WMAP) data are pointing towards the existence of additional light degrees of freedom $N_{\text{eff}} = 3.84 \pm 0.40$ \cite{wmap9}. Recently, the Planck collaboration have reported a preference towards the standard three light neutrino scenario $ N_{\text{eff}} = 3.30^{+0.54}_{-0.51}$ \cite{planck}. Nevertheless, the issue of more than three relativistic degrees of freedom is not yet settled. Apart from cosmological hints in support of light sterile neutrino, there have also been evidence from anomalous results in accelerator and reactor based neutrino experiments. The anomalous results in anti-neutrino flux measurements at the LSND accelerator experiment \cite{LSND1} provided the first hint of light sterile neutrinos. The LSND results have also gained support from the latest data released by the MiniBooNE experiment \cite{miniboone}. Similar anomalies have also been observed at nuclear reactor neutrino experiments \cite{react} as well as gallium solar neutrino experiments \cite{gall}. These anomalies suggesting the presence of light sterile neutrinos have led to global short-baseline neutrino oscillation data favoring two light sterile neutrinos within the eV range \cite{Kopp:2011qd}. Some more interesting discussions on light sterile neutrinos from cosmology as well as neutrino experiments point of view can be found in \cite{cosmo_ste} and references therein. Thus, the hints in favor of sub-eV scale sterile neutrinos have led to a model building challenge to explain the origin of three light active neutrinos together with one or two sterile neutrinos within the same mass range. Some interesting proposals along these lines have appeared recently in \cite{rodejohann,model}. A nice review of some of the earlier works can also be found in \cite{merle}.

In this letter, we present an abelian extension of the standard model where three active and one sterile neutrino masses arise at eV scale. The gauge charges of the field content under the additional $U(1)_X$ gauge group are chosen in such an anomaly free way that only one active neutrino acquires non-zero tree level mass from usual type I seesaw mechanism whereas two other active neutrinos and one sterile neutrino remain massless. However, at one-loop level two other active neutrinos and one sterile neutrino acquire non-zero mass. Due to the loop suppression, the particles in the loop can be around the TeV corner while keeping the neutrino masses at eV scale and hence can have interesting signatures in the colliders. This model also allows non-zero mixing between active and sterile neutrinos at one loop level and hence can have tantalizing consequences in the reactor neutrino experiments. Also, as discussed in one of our earlier works \cite{Borah:2012qr}, this model also has the provision of breaking the gauge symmetry spontaneously in a way that leaves a remnant $Z_2$ symmetry at low energy allowing the lightest $Z_2$ odd particle to be stable and hence a cold dark matter candidate. However, in our minimal setup, to allow non-trivial mixing of light sterile neutrino with the active neutrinos, we have to sacrifice this $Z_2$ symmetry. Thus, the present work is not aimed at explaining dark matter which may have origin from a different new physics sector.

This letter is organized as follows. In section \ref{model} we briefly discuss the model we are interested in. In section \ref{sec:numass} we study the generation of three active neutrino masses in this model. In section \ref{sec:stmass}, we discuss the origin of one eV scale sterile neutrino in our model. Then in section \ref{sec:acstmass} we discuss the possibility of active-sterile neutrino mixing at one-loop level and finally conlude in \ref{results}.

\section{The Model}
\label{model}
The model which we take as a starting point of our discussion was first proposed in \cite{Adhikari:2008uc}. The authors in that paper discussed various possible scenarios with different combinations of Majorana singlet fermions $N_R$ and Majorana triplet fermions $\Sigma_R$. Here we discuss one of such models which we find the most interesting for our purposes. This, so called model C by the authors in \cite{Adhikari:2008uc}, has the the following particle content shown in table \ref{table1}.

\begin{center}
\begin{table}
\caption{Particle Content of the Model}
\begin{tabular}{|c|c|c|c|}
\hline
Particle & $SU(3)_c \times SU(2)_L \times U(1)_Y$ & $U(1)_X$ & $Z_2$ \\
\hline
$ (u,d)_L $ & $(3,2,\frac{1}{6})$ & $n_1$ & + \\
$ u_R $ & $(\bar{3},1,\frac{2}{3})$ & $\frac{1}{4}(7 n_1 -3 n_4)$ & + \\
$ d_R $ & $(\bar{3},1,-\frac{1}{3})$ & $\frac{1}{4} (n_1 +3 n_4)$ & +\\
$ (\nu, e)_L $ & $(1,2,-\frac{1}{2})$ & $n_4$ & + \\
$e_R$ & $(1,1,-1)$ & $\frac{1}{4} (-9 n_1 +5 n_4)$ & + \\
\hline
$N_R$ & $(1,1,0)$ & $\frac{3}{8}(3n_1+n_4)$ & - \\
$\Sigma_{1R,2R} $ & $(1,3,0)$ & $\frac{3}{8}(3n_1+n_4)$ & - \\
$ S_{1R}$ & $(1,1,0)$ & $\frac{1}{4}(3n_1+n_4)$ & + \\
$ S_{2R}$ & $(1,1,0)$ & $-\frac{5}{8}(3n_1+n_4)$ & - \\
\hline
$ (\phi^+,\phi^0)_1 $ & $(1,2,-\frac{1}{2})$ & $\frac{3}{4}(n_1-n_4)$ & + \\
$ (\phi^+,\phi^0)_2 $ & $(1,2,-\frac{1}{2})$& $\frac{1}{4}(9n_1-n_4)$ & + \\
$(\phi^+,\phi^0)_3 $ & $(1,2,-\frac{1}{2})$& $\frac{1}{8}(9n_1-5n_4)$ & - \\
\hline
$ \chi_1 $ & $(1,1,0)$ & $-\frac{1}{2}(3n_1+n_4)$ & + \\
$ \chi_2 $ & $(1,1,0)$ & $-\frac{1}{4}(3n_1+n_4)$ & + \\
$ \chi_3 $ & $(1,1,0)$ & $-\frac{3}{8}(3n_1+n_4)$ & - \\
$ \chi_4 $ & $(1,1,0)$ & $-\frac{3}{4}(3n_1+n_4)$ & + \\
\hline
\end{tabular}
\label{table1}
\end{table}
\end{center}
The third column in table \ref{table1} shows the $U(1)_X$ quantum numbers of various fields which satisfy the anomaly matching conditions. The Higgs content chosen above is not arbitrary and is needed, which leads to the possibility of radiative neutrino masses in a manner proposed in \cite{Ma:2006km} as well as a remnant $Z_2$ symmetry. Two more singlets $S_{1R}, S_{2R}$ are required to be present to satisfy the anomaly matching conditions. In this model, the quarks couple to $\Phi_1$ and charged leptons to $\Phi_2$ whereas $(\nu, e)_L$ couples to $N_R, \Sigma_R$ through $\Phi_3$ and to $S_{1R}$ through $\Phi_1$. The extra four singlet scalars $\chi$ are needed to make sure that all the particles in the model acquire mass. The lagrangian which can be constructed from the above particle content has an automatic $Z_2$ symmetry and hence provides a cold dark matter candidate in terms of the lightest odd particle under this $Z_2$ symmetry. Part of the scalar potential of this model relevant for our future discussion can be written as
$$ V_s \supset \mu_1 \chi_1 \chi_2 \chi^{\dagger}_4 + \mu_2 \chi^2_2 \chi^{\dagger}_1 +\mu_3 \chi^2_3 \chi^{\dagger}_4 + \mu_4 \chi_1 \Phi^{\dagger}_1 \Phi_2 + \mu_5 \chi_3 \Phi^{\dagger}_3 \Phi_2 +\lambda_{13} (\Phi^{\dagger}_1 \Phi_1)(\Phi^{\dagger}_3 \Phi_3)$$
$$ +f_1\chi_1\chi^{\dagger}_2\chi^2_3+f_2\chi^3_2\chi^{\dagger}_4+f_3 \chi_1 \chi^{\dagger}_3\Phi^{\dagger}_1\Phi_3 +f_4 \chi^2_2\Phi^{\dagger}_1\Phi_2 + f_5 \chi^{\dagger}_3\chi_4 \Phi^{\dagger}_3 \Phi_2 $$
\begin{equation}
 +\lambda_{23} (\Phi^{\dagger}_2 \Phi_2)(\Phi^{\dagger}_3 \Phi_3)+ \lambda_{16} (\Phi^{\dagger}_1 \Phi_1)(\chi^{\dagger}_3 \chi_3) + \lambda_{26} (\Phi^{\dagger}_2 \Phi_2)(\chi^{\dagger}_3 \chi_3)
\label{scalpot}
\end{equation}

Let us denote the vacuum expectation values (vev) of various Higgs fields as $ \langle \phi^0_{1,2} \rangle = v_{1,2}, \; \langle \chi^0_{1,2,4} \rangle  =u_{1,2,4}$. We also denote the coupling constants of $SU(2)_L, U(1)_Y, U(1)_X$ as $g_2, g_1, g_x$ respectively. The charged weak bosons acquire mass $M^2_W = \frac{g^2_2}{2}(v^2_1+v^2_2) $. The neutral gauge boson masses in the $(W^{\mu}_3, Y^{\mu}, X^{\mu})$ basis is 
\begin{equation}
M =\frac{1}{2}
\left(\begin{array}{cccc}
\ g^2_2(v^2_1+v^2_2) & g_1g_2(v^2_1+v^2_2) &  M^2_{WX} \\
\ g_1g_2(v^2_1+v^2_2) & g^2_1(v^2_1+v^2_2) & M^2_{YX} \\
\ M^2_{WX} & M^2_{YX}  & M^2_{XX}
\end{array}\right)
\end{equation}
where 
$$M^2_{WX} = -g_2g_x(\frac{3}{4}(n_1-n_4)v^2_1+\frac{1}{4}(9n_1-n_4)v^2_2) $$
$$ M^2_{YX} = -g_1g_x(\frac{3}{4}(n_1-n_4)v^2_1+\frac{1}{4}(9n_1-n_4)v^2_2)$$
$$ M^2_{XX} = g^2_x(\frac{9}{4}(n_1-n_4)^2v^2_1+\frac{1}{4}(9n_1-n_4)^2v^2_2+\frac{1}{16}(3n_1+n_4)^2(4u^2_1+u^2_2+9u^2_4)) $$
The mixing between the electroweak gauge bosons and the additional $U(1)_X$ boson as evident from the above mass matrix should be very tiny so as to be in agreement with electroweak precision measurements. The stringent constraint on mixing can be avoided by assuming a very simplified framework where there is no mixing between the electroweak gauge bosons and the extra $U(1)_X$ boson. Therefore $ M^2_{WX} = M^2_{YX} = 0$ which gives rise to the following constraint
\begin{equation}
3(n_4-n_1)v^2_1 = (9n_1-n_4)v^2_2
\label{zeromixeq}
\end{equation}
which implies $1 < n_4/n_1 <9 $. If $U(1)_X$ boson is observed at LHC this ratio $n_4/n_1$ could be found empirically 
from its decay to $q\bar{q}$, $l\bar{l}$ and $\nu\bar{\nu}$ \cite{Adhikari:2008uc}. Here, $q$, $l$ and $\nu$ correspond to 
quarks, charged leptons and neutrinos respectively.
In terms of the charged weak boson mass, we have 
$$ v^2_1 = \frac{M^2_W(9n_1-n_4)}{g^2_2(3n_1+n_4)}, \quad v^2_2 = \frac{M^2_W(-3n_1+3n_4)}{g^2_2(3n_1+n_4)} $$
Assuming zero mixing, the neutral gauge bosons of the Standard Model have masses
$$ M_B = 0, \quad M^2_Z = \frac{(g^2_1+g^2_2)M^2_W}{g^2_2} $$
which corresponds to the photon and weak Z boson respectively. The $U(1)_X$ gauge boson mass is 
$$ M^2_X = 2g^2_X (-\frac{3M^2_W}{8g^2_2}(9n_1-n_4)(n_1-n_4)+\frac{1}{16}(3n_1+n_4)^2(4u^2_1+u^2_2+9u^2_4)) $$

\section{Active Neutrino Mass}
\label{sec:numass}

In this section, we summarize the origin of neutrino mass in the model. The relevant part of the Yukawa Lagrangian is 
$$ \mathcal{L}_Y \supset y \bar{L} \Phi^{\dagger}_1 S_{1R} + h_N \bar{L} \Phi^{\dagger}_3 N_R + h_{\Sigma}  \bar{L}\Phi^{\dagger}_3 \Sigma_R + f_N N_R N_R \chi_4+ f_S S_{1R} S_{1R} \chi_1 $$
\begin{equation}
+ f_{\Sigma} \Sigma_R \Sigma_R \chi_4 + f_{NS} N_R S_{2R} \chi^{\dagger}_2 + f_{12} S_{1R} S_{2R} \chi^{\dagger}_3
\label{yukawa} 
\end{equation}

\begin{figure}[htb]
\centering
\includegraphics[scale=0.75]{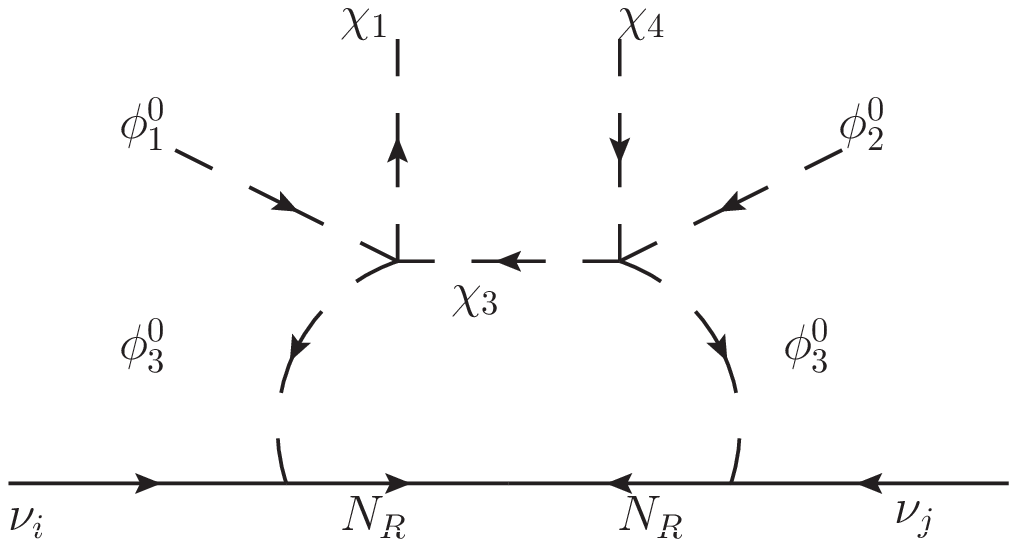}
\caption{One-loop contribution to active neutrino mass}
\label{numass}
\end{figure}

The Majorana mass of the fermions $S_R, N_R$ and $\Sigma_R$ arise as a result of the spontaneous symmetry breaking of $U(1)_X$ symmetry by the vev of $\chi_{1,2,4}$. Neutrinos acquire Dirac masses by virtue of their couplings to $S_{1R}$ as shown in equation $(\ref{yukawa})$. Thus the $3 \times 3$ neutrino mass matrix receives tree level contribution from standard type I seesaw mechanism and one gets the hierarchical pattern of neutrino mass with only one massive and other two neutrinos massless. $S_{1R}$ can couple to arbitrary linear combination of $\nu_i$ by assigning different values of $y$ in
equation (\ref{yukawa}) for
different generation. However, considering it non-zero  and same for $\nu_\mu$ and $\nu_\tau$ only,  the heaviest mass $m_{\nu 3}$ is 
given by 
\begin{equation}
m_{\nu 3} \approx \frac{ 2y^2 v_1^2}{f_S u_1}
\label{neutmass}
\end{equation}
which sets the scale of higher neutrino mass square difference of  about $ 2.4  \times 10^{-3}$ eV$^2$ and for that
$m_{\nu 3}  $ may be considered to be about $0.05$ eV (for hierarchical neutrino masses) to about 0.1 eV (for almoset degenerate neutrino masses).
Two neutrinos which are massless at the tree level become massive from one loop contribution 
as shown in figure \ref{numass} of Feynman diagram involving one of $N_R, \Sigma^0_{1R, 2R}$. These contributions 
set the scale of lower mass squared difference of about $7.6 \times 10^{-5}$ eV$^2$ for which other neutrino masses may be considered to be about $10^{-2}$ eV (for hierarchical neutrino masses ) to about 0.1 eV (for almost degenerate neutrino masses). Somewhat similar to \cite{Ma:2006km}, such a one loop diagram gives partial
contribution through $A_k$ as mentioned below when there is a mass splitting between the CP-even and CP-odd neutral components of the Higgs field involved in the loop which is $\phi^0_3$ in this case. In our considered model, such a mass splitting is possible due to the  couplings between $\phi^0_3$ and the singlet scalar fields $\chi$ shown in equation $(\ref{scalpot})$. Such a mass splitting is also necessary for $\phi^0_3$ to be a dark matter candidate as discussed in \cite{Borah:2012qr}.

The  one-loop contribution $(M_\nu)_{ij}$ to $3\times 3$ neutrino mass matrix is given by

\begin{eqnarray} 
({M_\nu)}_{ij} \approx  \frac{f_3 f_5 v_1 v_2 u_1 u_4}{16 \pi^2} \sum_k {h_{N, \Sigma} }_{ik} {h_{N, \Sigma} }_{jk} \left( A_k +{(B_k)}_{ij} \right)
\label{nuradmass}
\end{eqnarray}
where $k=1,2,3$ corresponds to different $N_R$,  
\begin{eqnarray}
A_k &=& {(M_{N, \Sigma})}_k \left[ I\left( m_{\phi^0_{3R}},m_{\phi^0_{3R}},{(M_{N, \Sigma})}_k, m_{\chi_{3R}} \right) - I\left(m_{\phi^0_{3I}},m_{\phi^0_{3I}},{(M_{N, \Sigma})}_k, m_{\chi_{3R}} \right) \right], 
\label{Ak}
\end{eqnarray}
\begin{eqnarray}
{(B_k)}_{ij}= -(2-\delta_{ij}) {(M_{N, \Sigma})}_k I\left(m_{\phi^0_{3R}},m_{\phi^0_{3I}},{(M_{N, \Sigma})}_k, m_{\chi_{3I}} \right),
\label{Bk}
\end{eqnarray}
in which
\begin{eqnarray}
I(a,a,b,c)=  \frac{(a^4-b^2 c^2) \ln (a^2/c^2)}{{(b^2 -a^2)}^2 {(c^2 -a^2)}^2}+ \frac{b^2 \ln (b^2/c^2)}{ (c^2 - b^2){(a^2 - b^2)}^2}  
  -\frac{1}{(a^2 - b^2) (a^2 -c^2)},
\end{eqnarray}
\begin{eqnarray} 
I(a,b,c,d)&=&  \frac{1}{a^2-b^2}  \left[  \frac{1}{a^2-c^2} \left( \frac{a^2}{a^2-d^2} \ln(a^2/d^2) - \frac{c^2}{c^2-d^2} \ln(c^2/d^2)\right)\right. \nonumber \\  &-&  \left. \frac{1}{b^2-c^2} \left( \frac{b^2}{b^2-d^2} \ln(b^2/d^2) - \frac{c^2}{c^2-d^2} \ln(c^2/d^2)\right)  \right]
\end{eqnarray}
and
$m_{\phi^0_{3R}}$ and $m_{\phi^0_{3I}} $ are the masses corresponding to $Re[\phi^0_3]$ and $Im[\phi^0_3]$ respectively
and  $m_{\chi_{3R}}$ and  $m_{\chi_{3I}}$ are the masses corresponding to $Re[\chi^0_3]$ and $Im[\chi^0_3]$ respectively  and 
$v_i= \langle \phi_i \rangle$ and $u_i= \langle \chi_i \rangle$ . 
$M_{N,\Sigma}$ is the Majorana mass term of $N_R(\Sigma^0_R)$. $h_{N, \Sigma}$ are the Yukawa couplings in equation $(\ref{yukawa})$. 
If $N_R$ is replaced by $\Sigma_R$ in the Feynman diagram then in the above expression $h_{\Sigma_{ij}}$ is to be considered instead of $h_{N_{ij}}$ as shown above. 

Under 
some assumptions $A_k$ and $B_k$ have simpler forms as given below. We
write  $(M_{N, \Sigma})_k$  as $m_{2k}$.
We have neglected the mixing between $\phi_3^0$ and $\chi_3^0$.
If all the scalar masses in the loop diagram are almost degenerate and written as $m_{sc}$ then 
\begin{eqnarray}
A_k + (B_k)_{ij} \approx m_{2k} \left[\frac{m_{sc}^2 
+ m_{2k}^2 }{m_{sc}^2 \left( m_{sc}^2 - m_{2k}^2 \right)^2 }- \frac{(2-\delta_{ij})\; m_{2k}^2}{\left(m_{sc}^2 - m_{2k}^2 \right)^3}\ln \left( m_{sc}^2/m_{2k}^2 \right)   \right],
\label{scaldeg}
\end{eqnarray}
and if all scalar and fermion masses in the loop are almost degenerate and written as $m_{deg}$ then
\begin{eqnarray}
A_k + (B_k)_{ij} \approx  \frac{(2-\delta_{ij})}{6 m_{deg}^3}\; .
\label{fermscal}
\end{eqnarray}
To get appropriate neutrino mass square differences we shall require these loop contributions to be about $10^{-2}$ eV 
(for hierarchical neutrino masses) to about $0.1$ eV {for almost degenerate neutrino masses) as mentioned earlier. 

\begin{figure}[htb]
\centering
\includegraphics[scale=0.75]{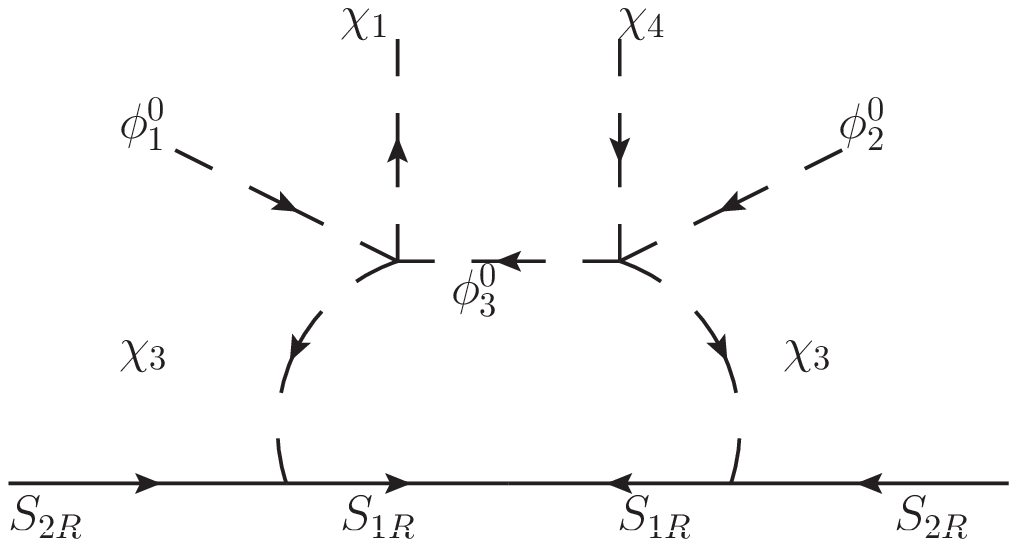}
\caption{One-loop contribution to sterile neutrino mass}
\label{sterile1}
\end{figure}
\begin{figure}[htb]
\centering
\includegraphics[scale=0.75]{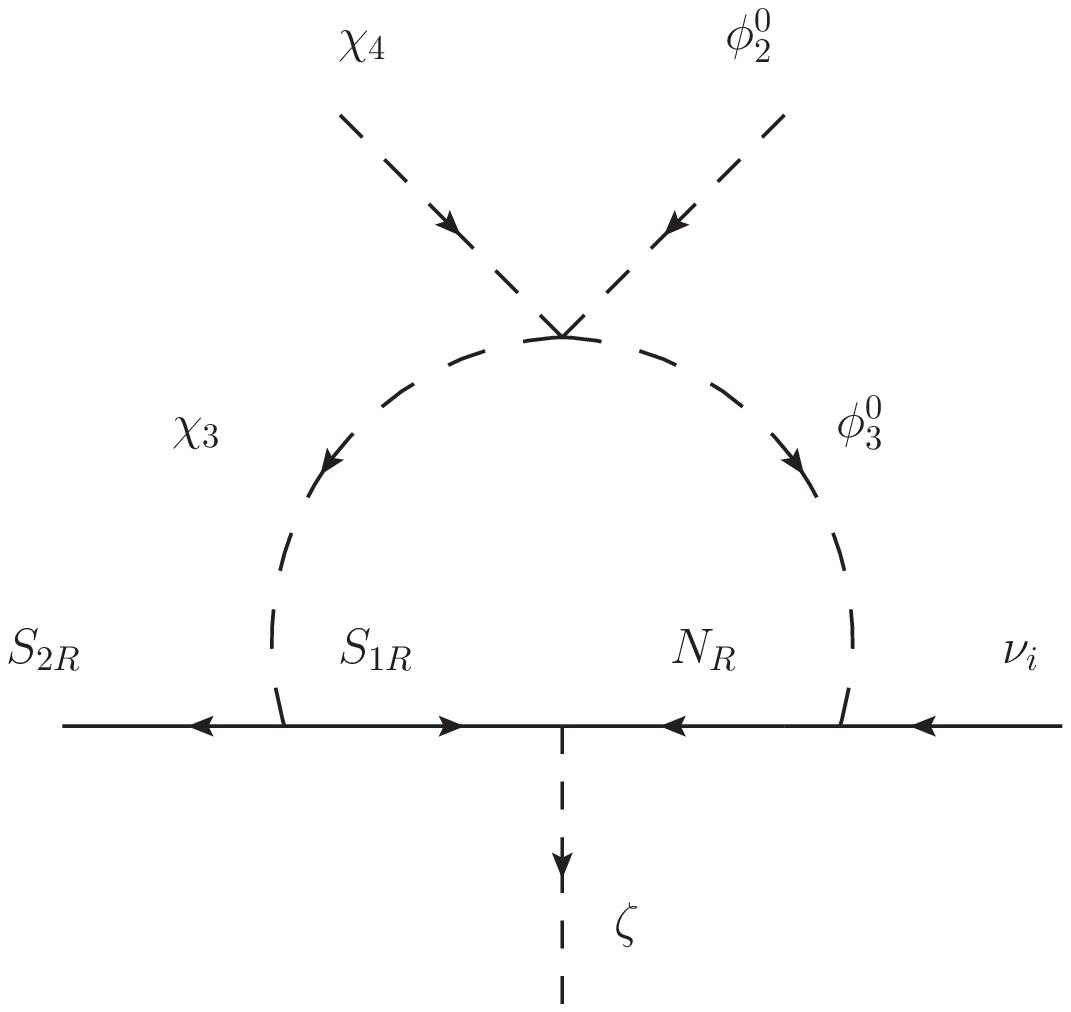}
\caption{One-loop contribution to active-sterile neutrino mixing}
\label{sterile2}
\end{figure}

After taking into account the one loop correction to tree level neutrino mass matrix (as shown in equation (\ref{numassmatrix}) below) we discuss about the mixing of different flavors of active neutrinos. As discussed earlier we consider $y_1 \approx 0, y_2 \approx y_3 \approx y$ in the $4 \times 4$ block of the neutrino mass matrix in (\ref{numassmatrix}) which give $m_{\nu 3}$ as shown in equation (\ref{neutmass}). This choice also gives maximal mixing of $\nu_\mu$ and $\nu_\tau$ at the tree level. As loop contributions to this block are relatively much smaller it is expected that they will not change this maximal mixing much which is required to support atmospheric neutrino oscillation data. For simplicity of the loop contribution let us assume $u_1 \sim u_4 \sim m_{deg}$
and $f_3 \sim f_5 \sim 10^{-3}$, without losing any generality. 
Considering equation (\ref{nuradmass}) and (\ref{fermscal}) we define
$$
a_i=  \left\{\frac{f_3 f_5 v_1 v_2 u_1 u_4}{16 \pi^2}   \frac{(2-\delta_{ij})}{6 m_{deg}^3}\right\}^{1/2}\sum_k {h_{N, \Sigma} }_{ik} .
$$
(where all ${h_{N, \Sigma} }_{ik}$ are assumed to be equal for particular $i$ value) such that we can write the one loop contribution to $3\times 3$ active neutrino mass matrix as $(M_{\nu})_{ij} \approx a_i a_j$ where $i=1,2,3$.  Let us consider ${h_{N, \Sigma} }_{ik} \sim 10^{-3}$ for $i=2,3$. This implies $a_2 \sim a_3$ which are written as $a$ here onwards. These choices give $m_{\nu 2} \sim a^2 \sim 10^{-2} $ eV  which is, for example, appropriate for hierarchical neutrino masses. 

After diagonalizaion of tree level neutrino mass matrix although there is rotation in the $\nu_\mu$ and $\nu_\tau$ basis, the order of different matrix elements will not change from the order of the loop contributions in the upper left $3 \times  3$ block except the $33$ element. The $33$ element is given by $m_{\nu 3}$ which is about one order of magnitude higher than other elements. Therefore, corresponding to $\nu_e - \nu_\tau$ mixing the angle $\theta_{13}$ will be naturally small and is given by
$$ \tan 2 \theta_{13} \approx 2 a_1 a/m_{\nu 3}\; $$
This could be about $0.3$ for our choices of parameters (with $a_1 \approx 0.08$  ) resulting in $\sin ^2 2 \theta_{13}  \approx 0.1$ as required by recent experiments like Daya Bay. Corresponding to $\nu_e - \nu_\mu$ mixing, the angle $\theta_{12}$ is given by
$$ \tan 2 \theta_{12}  \approx 2 a_1 a /(a^2 -a_1^2)\; .$$  Considering 
${h_{N, \Sigma} }_{1k}$ somewhat close to ${h_{N, \Sigma} }_{ik}$ with $i=2,3$ as required by our above-mentioned choices of parameters, it is possible to obtain nearly maximal value of $\theta_{12}$ as obtained from solar neutrino oscillation data.
\section{Sterile Neutrino Mass}
\label{sec:stmass}
In the model we are studying, there are three singlet fermions $S_{1R}, S_{2R}, N_R$. Out of these, $S_{1R}$ couples to the active neutrinos and contribute to tree level mass term of one of the active neutrinos as discussed in the previous section. Thus, for generic Dirac Yukawa couplings of the neutrinos, the sterile neutrino $S_{1R}$ is expected to be much heavier than the eV scale. Otherwise, one of the active neutrinos will receive a large tree level contribution to its mass in disagreement with observations.  Thus either $S_{2R}$ or $N_R$ or both could give rise to the light sterile neutrinos. From the Yukawa lagrangian \ref{yukawa}, we see that there is a tree level mass term $f_N \langle \chi_4 \rangle$ which is
 certainly not of eV order as we are considering the additional $U(1)_X$ symmetry to be broken (by the vev of $\chi$) at a scale above the electroweak scale. So $N_R$ is not a preferred candidate for light sterile neutrino in our model. However, there is no tree level mass term for $S_{2R}$. There are mixing terms of $S_{2R}$ with $N_R, S_{1R}$ through Higgs fields $\chi_2, \chi_3$ respectively. Out of these two, the field $\chi_3$ does not acquire vev owing to the fact that it is $Z_2$-odd. If the vev of the other Higgs field $\chi_2$ is zero, the singlet fermion $S_{2R}$ decouples from rest of the fermions at tree level. As seen from the discussion on active neutrino masses, it can be noted that the vev of $\chi_2$ does not appear in the mass formula \ref{nuradmass} and hence can safely be turned off. In our notation, the full fermion mass matrix at tree level looks like the one below
\begin{equation}
M_f =
\left(\begin{array}{cccccc}
\ 0 & 0 & 0 & y_1v_1 & 0 & 0 \\
\ 0 & 0 & 0 & y_2v_1 & 0 & 0 \\
\ 0 & 0 & 0 & y_3 v_1 & 0 & 0 \\
\ y_1v_1 & y_2v_1 & y_3v_1 & f_S u_1 & 0 & 0 \\
\ 0 & 0 & 0 & 0 & 0 & 0 \\
\ 0 & 0 & 0 & 0 & 0 & f_N u_4
\end{array}\right)
\label{numassmatrix}
\end{equation}
The above mass matrix is written in the basis $(\nu_e, \nu_{\mu}, \nu_{\tau}, S_{1R}, S_{2R}, N_R)$. Thus, the sterile neutrino $S_{2R}$ remains massless at tree level and can aquire a small eV scale mass at one loop level from the diagram \ref{sterile1}. Since $S_{2R}$ corresponds to the fifth entry in the mass matrix above, we denote its mass term as $(M_f)_{55}$. The one-loop contribution to its mass can be written as
\begin{eqnarray} 
{(M_f)}_{55} \approx  \frac{f_{12}^2 f_3 f_5 v_1 v_2 u_1 u_4}{16 \pi^2}  \left( A +B \right)
\label{nuradmass1}
\end{eqnarray}
where $A$ and $B$ can be obtained by replacing ${(M_{N, \Sigma})}_k$ by $M_{S_{1R}}$ in $A_k$ and $B_k$ in $(\ref{Ak})$ and $(\ref{Bk})$ respectively.

\section{Active Sterile Neutrino Mixing}
\label{sec:acstmass}
As discussed in the last section, a light eV scale sterile neutrino can be naturally accommodated in our model. It has zero tree level mass but obtains a mass of the same scale as the light active neutrinos through one loop corrections. However, to have implications in neutrino oscillation experiments, this light sterile neutrino should have non-trivial mixing with the active neutrinos. It can be seen from our discussion above that the light sterile neutrino $S_{2R}$ does not have any tree level mixing with the active neutrinos. Even at loop level, there is no active sterile neutrino mixing with the minimal field content discussed above. We introduce a new scalar singlet field $\zeta$ with $U(1)_X$ charge $\frac{5}{8}(3n_1+n_4)$ which a tree level mixing term of $N_R$ and $S_{1R}$ which effectively allow one loop mixing between active and sterile neutrino as seen in the diagram \ref{sterile2}. The mixing term corresponding to this diagram can be estimated as
\begin{eqnarray} 
{(M_f)}_{5j} = {(M_f)}_{j5}^* \approx  \sum_k \frac{f_{12} f_5  v_2 u_4 (h_{N, \Sigma })_{kj} M_X }{16 \pi^2} \left[  I\left(m_{\chi_{3R}},m_{\phi_{3R}^0}, M_X\right)-I\left(m_{\chi_{3I}},m_{\phi_{3I}^0}, M_X\right)\right]  \nonumber \\
\label{nuradmass2}
\end{eqnarray}
where $j, k = 1,2,3$ and $k$ corresponds to different $N_R$.
$M_X$ is the $U(1)_X$ symmetry breaking scale and we have assumed 
$ M_{S_{1R}} \sim M_{N_R}\sim M_{\Sigma_R} \sim M_X $ and 
\begin{eqnarray} 
I\left( a,b,c\right) =  \frac{a^2 b^2 \ln(a^2/b^2) + b^2 c^2 \ln(b^2/c^2)+c^2 a^2 \ln(c^2/a^2)}{(a^2-b^2) (b^2 -c^2) (c^2-a^2)}
\end{eqnarray}
If $N_R$ is replaced by $\Sigma_R$ in the Feynman diagram then in the above expression $h_{\Sigma_{kj}}$ is to be considered instead of $h_{N_{kj}}$ as shown above. If we consider   $m_{\chi_{3R}} \sim m_{\phi_{3R}^0}$ and/or $m_{\chi_{3I}} \sim m_{\phi_{3I}^0}$ then to get ${(M_f)}_{5j} $
one is required to consider $I\left( a,a,c\right)$ which is given by 
\begin{eqnarray} 
I\left( a,a,c\right) =  \frac{1}{{(a^2-c^2)}^2}\left( a^2-c^2 - c^2 \ln (a^2/c^2) \right)
\end{eqnarray}
Assuming $m_{\chi_{3R}} \sim m_{\chi_{3I}} \ll M_X$, we have 
\begin{equation} 
{(M_f)}_{5j}\approx  \sum_k \frac{f_{12} f_5  v_2 u_4 (h_{N, \Sigma })_{kj} M_X }{16 \pi^2 M^4_X} \left[ m^2_{\chi_{3R}}-m^2_{\chi_{3I}}-M^2_X \ln ( m^2_{\chi_{3R}}/m^2_{\chi_{3I}})\right]
\end{equation}
Writing $m_{\chi_3}= (m_{\chi_{3R}}+m_{\chi_{3I}})/2$ and assuming the Yukawa couplings to be real 
\begin{equation}
{(M_f)}_{15}\approx  \sum_k \frac{f_{12} f_5  v_2 u_4 (h_{N, \Sigma })_{1k} }{16 \pi^2 M_X m_{\chi_3}^2 } \left[ m^2_{\chi_{3R}}-m^2_{\chi_{3I}}\right]
\end{equation} 
For ${(M_f)}_{25}$ in the above expression $(h_{N, \Sigma })_{1k}$ will be replaced by $(h_{N, \Sigma })_{2k}$.  As 
$A+B \approx 1/( m_{\chi_3}^2 M_X)$ 
\begin{eqnarray} 
{(M_f)}_{55} \approx  \frac{f_{12}^2 f_3 f_5 v_1 v_2 u_1 u_4}{16 \pi^2 m_{\chi_3}^2 M_X }
\end{eqnarray}
If we consider $(h_{N, \Sigma })_{1k}^2 \sim (h_{N, \Sigma })_{2k}^2 \ll f_{12}^2$ then   ${(M_f)}_{11} \sim {(M_f)}_{22} \ll {(M_f)}_{55}$ and we can write the active sterile mixing angles as 
\begin{equation}
\tan 2\theta_{e5} = \frac{2{(M_f)}_{15}}{{(M_f)}_{55}} ;\;\;\; \tan 2\theta_{\mu 5} = \frac{2{(M_f)}_{25}}{{(M_f)}_{55}} 
\end{equation}
and these active sterile mixing angles could be small. However, if we consider
$(h_{N, \Sigma })_{1k}^2 \sim (h_{N, \Sigma })_{2k}^2 \sim f_{12}^2$ then   ${(M_f)}_{11} \sim {(M_f)}_{22} \sim {(M_f)}_{55}$ and these mixing angles could be large which
is not expected.

One may note that unlike the conventional 4th row and column for the sterile entry in the neutrino mass matrix, we have written it as the fifth entry as can be seen in the tree level neutrino mass matrix (\ref{numassmatrix}). The fourth entry corresponds to the heavy sterile neutrino $S_{1R}$. To compare with the global fit data for $3+1$ sub-eV neutrino scenario, we now go back to the conventional notation and denote the light sterile neutrino as the fourth entry and the heavy sterile neutrino as the fifth. As $\theta_{e5}$ and $\theta_{\mu 5}$ are very small, we can consider the $(1,4)$ element of the mixing matrix as
\begin{equation}
|U_{e4}| \sim  \sin\theta_{e5} \approx \frac{1}{2} \tan 2\theta_{e5} = \sum_k \frac{ (h_{N, \Sigma })_{1k} }{f_{12} f_3  v_1 u_1 } \left[ m^2_{\chi_{3R}}-m^2_{\chi_{3I}}\right] 
\label{sterilemix}
\end{equation}
To get $|U_{\mu 4}|$ in the above expression $(h_{N, \Sigma })_{1k}$ is to be replaced by $(h_{N, \Sigma })_{2k}$. 
The global neutrino fit data with three active and one light sterile neutrinos \cite{sterileglobal} give the best fit parameters as $\Delta m^2_{41} = 0.93 \; \text{eV}^2, \; \lvert U_{e4} \rvert = 0.15, \; \lvert U_{\mu 4} \rvert  = 0.17$ which, following the discussion above, can naturally be explained within the framework of our model. As for example, considering $f_{12} \approx 10^{-5}$ and other Yukawa couplings of the order of $10^{-6}$   and mass splitting between the real and imaginary component of the fields $\chi_3, \phi^0_3$ such that $(m^2_{\chi_{3R}}-m^2_{\chi_{3I}}) \approx 0.1 \; \text{GeV}^2 $ and the $vev$'s $v_1 \approx 100 \; \text{GeV}, \;u_1 \approx 1 \; \text{TeV}$ give the required order for active-sterile mixing.  In a less fine-tuned scenario, one can also have $f_{12} \approx 10^{-2}$, other Yukawa couplings of the order of $10^{-3}$ and the mass splitting approximately $100 \; \text{GeV}^2$ such that the desired mixing can be obtained. Using the experimental data it turns out
\begin{equation}
\sum_k \frac{(h_{N, \Sigma })_{1k}}{(h_{N, \Sigma })_{2k}} \approx  \frac{|U_{e4}| }{|U_{\mu 4}| }  \approx \frac{0.15}{0.17}
\end{equation}
and as such the Yukawa couplings $ (h_{N, \Sigma })_{1k} $ and  $(h_{N, \Sigma })_{2k}$ may be almost equal in our model.

While using the global fit data \cite{sterileglobal} for sterile neutrinos to constrain our model parameters, we note that these global fit data suffer from the tension between antineutrino appearance signals (as seen in accelerator, reactor and gallium anomalies) and bounds from antineutrino disappearance experiments \cite{disappear}. As concluded in \cite{sterileglobal}, a consistent interpretation of all data suggesting eV scale sterile neutrino is still missing and hopefully future oscillation and cosmology experiments will shed more light into it.
\section{Results and Conclusion}
\label{results}
Motivated by the hints from cosmology as well as neutrino experiments favoring additional light degrees of freedom, here we have studied an abelian extension of the standard model where three active and one sterile neutrino can acquire masses at the electron volt scale. At tree level, only one active neutrino acquires masses through type I seesaw mechanism whereas two active and one sterile neutrino remain massless. At one loop level, all the three active and one sterile neutrino receives non-zero mass contribution. Due to the radiative origin of the light neutrino masses, the new physics in this model can lie well around the TeV scale and hence can have interesting signatures in the collider experiments. This model also allows non-zero mixing between active and sterile neutrinos at loop level and hence can provide a suitable explanation to reactor neutrino anomalies. Such a mixing can also have non-trivial consequences in neutrinoless double beta decay experiments as recently discussed in \cite{rodejohann,sterileNDBD}.

It should be noted that the $Z_2$-odd singlet fermion $S_{2R}$ is being proposed as the sub-eV sterile neutrino in our work. Hence the usual weakly interacting massive particle (WIMP) dark matter scenario is lost in our framework. However, it is possible to project $S_{2R}$ as warm dark matter candidate by suitably adjusting the radiative mass to be in the keV region without any mixing with the light neutrino sector. We leave such a study for future investigation.

\section{Acknowledgement}
DB would like to acknowledge the hospitality at Center for Theoretical Physics, Jamia during January, 2013 where part of this work was completed.


\end{document}